\documentclass{PoS}

\title{Dark Stars: D\"od och \AA teruppst\aa ndelse}

\ShortTitle{Dark Stars: D\"od och \AA teruppst\aa ndelse}

\author{\speaker{Douglas Spolyar}$^1$,
Katherine Freese$^2$, 
Paolo Gondolo$^3$, 
Anthony Aguirre$^1$, 
Peter Bodenheimer$^1$, 
Jeremy A. Sellwood$^4$, and
Naoki Yoshida$^5$\\
        $^1$University of California, Santa Cruz, $^2$University of Michigan, $^3$University of Utah, $^4$Rutgers University, $^5$University of Tokyo\\
        E-mail: \email{dspolyar@physics.ucsc.edu}, \email{ktfreese@umich.edu}, \email{paolo@physics.utah.edu}, \email{aguirre@scipp.ucsc.edu}, \email{peter@ucolick.org}, \email{sellwood@physics.rutgers.edu}, \email{nyoshida@a.phys.nagoya-u-ac.jp} }

\abstract{The first phase of stellar evolution in the history of the universe
may be Dark Stars, powered by dark matter heating rather than by fusion.
Weakly interacting massive particles, which are their own
antiparticles, can annihilate and provide an important heat source for
the first stars in the universe.  This and the previous contribution present the story of Dark Stars. In this second part, we describe the structure of Dark Stars and predict that they are
very massive ($\sim 800 M_\odot$), cool (6000 K), bright ($\sim 10^6 L_\odot$), 
long-lived ($\sim 10^6$ years), and probable precursors to (otherwise
unexplained) supermassive black holes. 
Later, once the initial dark matter fuel runs out
and fusion sets in, dark matter annihilation can predominate again if the scattering
cross section is strong enough, so that a Dark Star is born again.
}

\FullConference{Identification of dark matter 2008\\
		 August 18-22, 2008\\
		 Stockholm, Sweden}

\begin{document}

\section{Building up the Mass}

Recently, we have found the stellar structure
of the dark stars (hereafter DS) 
(\cite{FreeseBodenheimerSpolyarGondolo08}).  
Though they form with the properties mentioned in the accompanying contribution (\cite{DarkStarsI}),
they continue to accrete mass from the surrounding medium. 
In (\cite{FreeseBodenheimerSpolyarGondolo08}) we build up the DS
mass as it grows from $\sim 1 M_\odot$
to $\sim 1000 M_\odot$.  
As the mass increases, the DS contracts and the DM
density increases until the DM heating matches its radiated luminosity.  We find
polytropic solutions for dark stars in hydrostatic and thermal
equilibrium.  We start
with a few $M_\odot$ DS and find an equilibrium solution.  Then we build up
the DS by accreting $1 M_\odot$ at a time with an accretion rate
of $2 \times 10^{-3} M_\odot$/yr, always finding equilibrium
solutions.  We find that initially
the DS are in convective equilibrium; from $(100-400)
M_\odot$ there is a transition to radiative; and heavier
DS are radiative.  As the DS grows, it pulls in more
DM, which then annihilates.  We continue this process until
the DM fuel runs out at $M_{DS} \sim 800 M_\odot$ (for 100 GeV WIMPs). 
Figure 3 shows the stellar structure. One can see
``the power of darkness:'' although the DM constitutes a tiny fraction
($<10^{-3}$) of the mass of the DS, it can power the star. The reason
is that WIMP annihilation is a very efficient power source:
2/3 of the initial energy of the WIMPs is converted into useful
energy for the star, whereas only 1\% of baryonic rest mass energy
is useful to a star via fusion.

\begin{figure}[b]
\centerline{\includegraphics[width=0.5\textwidth]{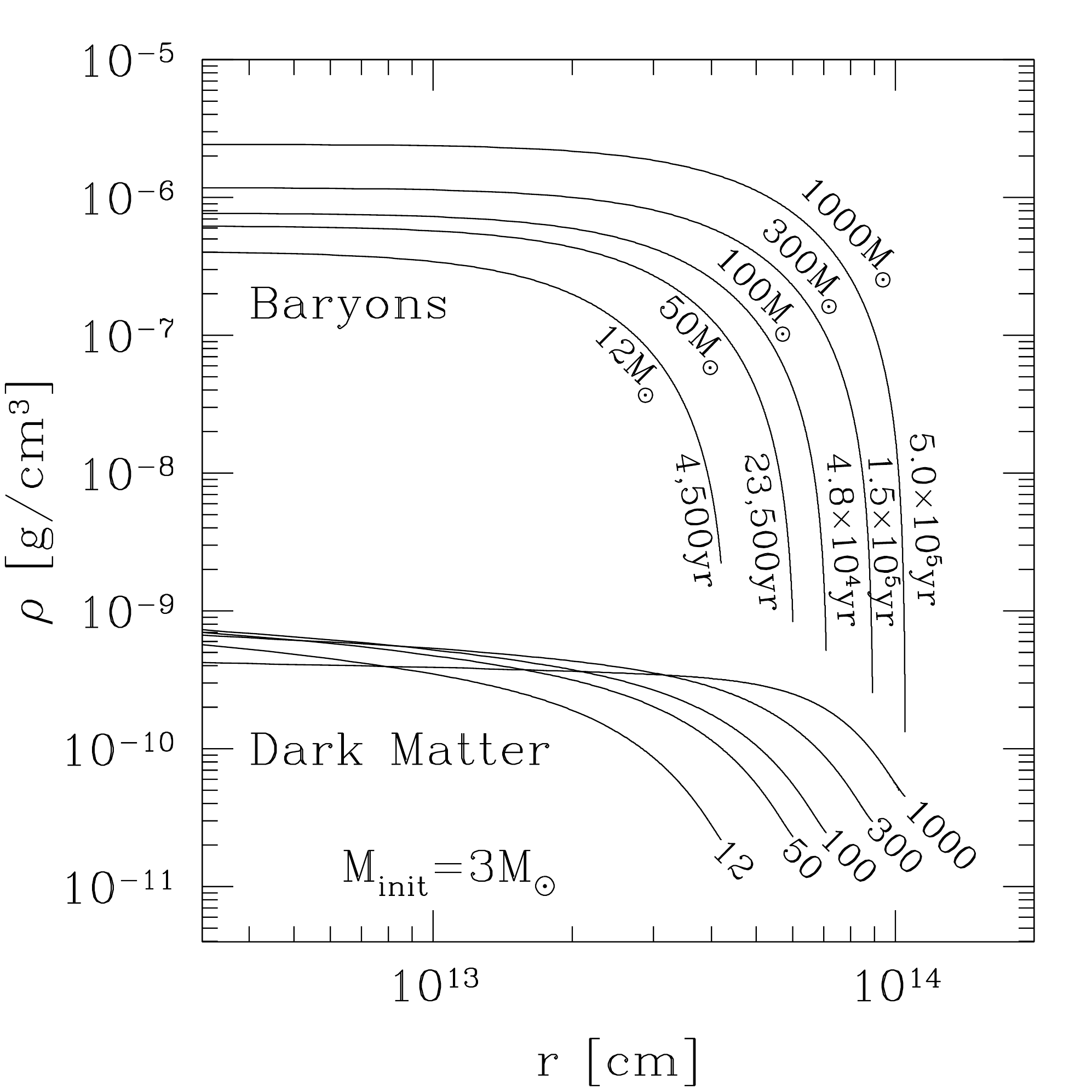}}
\caption{ Evolution of a dark star ($n$=1.5) as mass is accreted onto the initial
protostellar core of 3 M$_\odot$.  The set of upper
(lower) curves correspond to the baryonic (DM) density profile at different
masses and times.  Note that DM constitutes $<10^{-3}$ of the mass of the DS.
\vspace{-0.5\baselineskip}
}
\end{figure}

\section{Results and Predictions}

 Our final result 
(\cite{FreeseBodenheimerSpolyarGondolo08}), 
is very large first stars.
For example, for 100 GeV WIMPs, the first stars have $M_{DS} = 800 M_\odot$.
Once the DM fuel runs out inside the DS, the star contracts until
the central temperature reaches $10^8$ K and fusion sets in.
A possible end result of stellar evolution will be large black holes. 
The Pair Instability
SN (\cite{HegerWoosley02})
that would be produced from 140-260 $M_\odot$ stars (and whose
chemical imprint is not seen) would not be as abundant.  Indeed 
this process may help to explain the supermassive black
holes that have been found at high redshift ($10^9 M_\odot$ BH at
z=6) and are, as yet, unexplained 
(\cite{Li_etal07};
\cite{Pelupessy_etal07}). 

The lifetime of this new DM powered phase of stellar evolution, 
prior to the onset of fusion,
is $\sim 10^6$ years. The stars are very bright,
$\sim 10^6 L_\odot$, and relatively cool, (6000-10,000)K (as opposed
to standard Pop III stars whose surface temperatures exceed $30,000K$).  
Reionization
during this period is likely to be slowed down, as these stars can
heat the surroundings but not ionize them.  
One can thus
hope to find DS and differentiate them from standard Pop III stars;
perhaps some even still exist to low redshifts.

\section{Later stages: Capture}

There is another possible source of DM in the first stars: capture
of DM particles from the ambient medium. Whereas capture is negligible
during the pre-mainsequence phase, once fusion sets in it can be important,
depending on the value of the scattering cross section of DM with the gas.
Two simultaneous papers
(\cite{FreeseSpolyarAguirre08,Iocco08})
found the same basic idea: the DM luminosity from captured WIMPs
can be larger than fusion for the DS. Two uncertainties exist here:
the scattering cross section, and the amount of DM in the ambient medium
to capture from\footnote{Unlike the annihilation cross section, 
which is set by the relic
density, scattering is to some extent a free parameter set only
by bounds from direct detection experiments.}.
DS studies including capture
have assumed the maximal scattering cross sections allowed by
experimental bounds and ambient DM densities that are never depleted.
With these assumptions, 
DS evolution models with DM heating after the onset of fusion
have now been studied in several papers
(\cite{Iocco_etal08,Taoso_etal08,Yoon_etal08}).  We have been pursuing similar research
with Alex Heger on DS  
evolution after the onset of fusion.  

\section{Conclusion}

The line of research begun in Florence over two
years ago is reaching a very fruitful stage of development.
Dark matter can play a crucial role in the first stars.  The first
stars to form in the universe may be Dark Stars: powered by DM heating
rather than by fusion.  Our work indicates that they may 
be very large ($850 M_\odot$ for 100 GeV mass WIMPs).
The connections between particle physics
and astrophysics are ever growing!

\end{document}